\documentclass{JHEP}

\font\mybb=msbm10 at 12pt
\def\bb#1{\hbox{\mybb#1}}
\def\Z {\bb{Z}}

\title{On the Structure of Noncommutative N=2 Super Yang--Mills Theory}

\author{Diego BELLISAI, Jos\'e M. ISIDRO and Marco MATONE 
\\ Dipartimento di Fisica 
``G. Galilei'' -- Istituto Nazionale di Fisica Nucleare\\ Universit\`a di 
Padova, Via Marzolo, 8 -- 35131 Padova, Italy
\email{bellisai@pd.infn.it, isidro@pd.infn.it, matone@pd.infn.it}}

\abstract{We show that the recently proposed formulation
of noncommutative $N=2$ Super Yang--Mills theory implies that the 
commutative and noncommutative effective coupling constants 
$\tau(u)$ and $\tau_{nc}(u)$ coincide. We then introduce a key 
relation which allows to find a nontrivial solution of such 
equation, thus fixing the form of the low--energy effective action. 
The dependence on the noncommutative parameter arises from 
a rational function deforming the Seiberg--Witten differential.}

\preprint{DFPD00/TH/27\\{\tt hep-th/0009174}}


\begin{document}

\noindent
Noncommutative string and gauge theories have attracted strong 
attention \cite{CONNES,SW2,DH}. It is well known that gauge theories 
on a noncommutative space--time can arise as the low--energy 
effective open string theory in the presence of D--branes with a 
non--vanishing NS--NS two--form 
$B$--field \cite{SW2,DH,SH}.
An interesting related investigation concerns the formulation of the 
noncommutative version of $N=2$ Super Yang--Mills theory 
with gauge group $U(2)$ \cite{SHEIKH,Yoshida}. 

In this letter we argue that the deformation induced by the 
space--time noncommutativity can be neatly reabsorbed into a 
redefinition of the electric and magnetic masses $a$ and $a_D$ 
appearing in the BPS mass formula. In particular, we will derive an 
explicit expression for $a_{D,nc}$ and
$a_{nc}$ which denote the noncommutative analogues of $a_D$ and $a$.

In \cite{Yoshida} it has been found that, under reasonable assumptions, 
$a_{D,nc}$ and $a_{nc}$ have the same monodromies as their commutative 
partners \cite{SW}. Furthermore, the same elliptic curve that 
first appeared in
\cite{SW} has been found to describe the noncommutative theory.
The asymptotic behavior at $u=\infty$ is the same as in the commutative 
Seiberg--Witten model, {\it i.e.}
\begin{equation}
a_{D,nc}(u\to\infty)\sim \frac{i}{\pi}\sqrt{2u}\ln
\frac{u}{\Lambda^2}\ \ ,
\qquad a_{nc}(u\to\infty)\sim 
\sqrt{2u}\ \ .
\label{pocjw}\end{equation}
However, the asymptotic behavior of $a$ and $a_D$ in the dual 
$U(1)$ phase 
differs from its commutative counterpart, since the 
$\beta$--function gets 
also a contribution from the 
$U(1)$ gauge multiplet, which renders this theory
asymptotically free
\cite{SEI}. In fact, at $u=\Lambda^2$ we have  
\begin{equation}
a_{D,nc}(u\to\Lambda^2)\sim c_0(u-\Lambda^2)^{-1}\ \ ,
\label{d0idj}\end{equation} 
which has to be compared with the commutative case
\begin{equation}
a_{D}(u\to\Lambda^2)\sim \frac{i}{2\Lambda}(u-\Lambda^2)\ \ .
\label{anfamone}\end{equation}
Following these assumptions, in this letter we propose a definition 
of $a_{nc}$ and $a_{D,nc}$ through  a simple modification of the
Seiberg--Witten differential, and therefore of 
$a$ and $a_D$, which provides them with the 
same monodromies and asymptotic properties of $a_{nc}$ and $a_{D,nc}$.  

The framework of the derivation is similar to the one used in
\cite{M3} to prove the uniqueness of the Seiberg--Witten solution
by means of reflection symmetry of the quantum vacua.

According to \cite{Yoshida}, the behavior of the noncommutative 
effective gauge coupling constant $\tau_{nc}$ (as a function of $u$) 
for $u\to\infty$ and 
$u=+\Lambda^2$ is the same of $\tau$. 
Furthermore, since $a_{nc}$ and $a_{D,nc}$ have the same monodromy of
$a_D$ and $a$, it follows that $\tau_{nc}$ has 
the same monodromy of $\tau$. A further physical requirement on 
$\tau_{nc}$ is the positivity of its imaginary part
\begin{equation}
{\rm Im}\,\tau_{nc}= {4\pi\over g^2}>0\ \ .
\label{wdxow}\end{equation}
On the other hand, we know that the $u$ moduli space is 
the thrice punctured Riemann sphere. Thus, on general grounds, we 
can use the standard arguments of the uniformization theory, 
concerning the uniqueness of the uniformizing map \cite{M12,M3}, to 
see that 
\begin{equation}
\tau_{nc}(u)=\tau(u)\ \ .
\label{poojq}\end{equation}
This is a key point since it will lead us to fix the (polymorphic) 
functions $a_{nc}$ and $a_{D,nc}$. Actually, we will present an 
argument, which is in fact of interest also in uniformization 
theory, which will lead us to find a nontrivial solution to the 
following question. While on one side we have 
$\tau_{nc}(u)=\tau(u)$, on the other side we have that 
$a_{nc}$ and $a_{D,nc}$ do not coincide with
$a$ and $a_{D}$. Thus we are led to formulate the following problem:

\vspace{.333cm}

\noindent
{\it Given two sets of polymorphic functions $(a_{D,nc},a_{nc})$ and
$(a_D,a)$, having the same monodromy transformations,
find nontrivial solutions of the equation (\ref{poojq}), that is}

\begin{equation}
{\partial_u a_{D,nc}\over \partial_u a_{nc}}= {\partial_u a_{D}\over 
\partial_u a}\ \ .
\label{razzo}\end{equation}

\vspace{0.2cm}

\noindent
Since $a_{nc}$ and $a_{D,nc}$ have the same monodromies as $a$ and 
$a_{D}$, it would seem at first sight that 
$(a_{D,nc},a_{nc})=h(u)(a_D,a)$, where 
$h$ is a function of $u$ with trivial monodromies. However, this
would not solve Eq.(\ref{razzo}), unless $h=cnst$. Since from 
(\ref{d0idj}) and (\ref{anfamone}) we have 
$(a_{D,nc},a_{nc})\not\propto(a_D,a)$, it is clear that we
have to look for other functions. This is an important point because 
the proposal in \cite{Yoshida} may be implemented only if 
(\ref{razzo}) admits nontrivial solutions. It is remarkable that 
these solutions indeed exist. 
Let us start by recalling the differential equation \cite{LERCHE,M12}
\begin{equation}
\left(\partial_u^2+\frac{1}{4(u^2-\Lambda^4)}\right)
\pmatrix{a_D\cr a}=0\ \ .
\label{embhe}\end{equation}
We then consider two functions
$f(u)$ and $g(u)$ with trivial monodromy around $u=\infty$,
$u=\pm\Lambda^2$, and set
\begin{equation}
a_{D,nc}=f(u)a_D+g(u)a_D'\ \ ,\qquad a_{nc}=f(u)a+g(u)a'\ \ ,
\label{wodihw}\end{equation}
where $'\equiv\partial_u$. Note that $(a_{D,nc},a_{nc})$ in 
(\ref{wodihw}) have the same monodromy of $(a_{D},a)$, {\it i.e.} 
\begin{equation}
\pmatrix{a_{D,nc}\cr a_{nc}}\ \longrightarrow\  
\pmatrix{\tilde a_{D,nc}\cr \tilde a_{nc}}=M 
\pmatrix{a_{D,nc}\cr a_{nc}}\ \ ,
\end{equation}
where 
\begin{equation}
M_{\infty}=\pmatrix{-1&2\cr 0&-1}\ \ ,\qquad
M_{+1}=\pmatrix{1&0\cr -2&1}\ \ ,\qquad 
M_{-1}=\pmatrix{-1&2\cr -2&3}\ \ ,
\end{equation}
are the monodromies around $u=\infty,+\Lambda^2$ and 
$-\Lambda^2$ respectively. 

The crucial observation is that  
$a'_{D,nc}$ and $a'_{nc}$ still have the same form of
(\ref{wodihw}) with new functions $\tilde f$ and $\tilde g$. 
Actually, from (\ref{embhe}) and (\ref{wodihw}) we have 
\begin{equation}
a_{D,nc}'=\tilde f(u)a_D+\tilde g(u)a_D'\ \ ,\qquad a'_{nc}=
\tilde f(u)a+\tilde g(u)a'\ \ ,
\label{wodihw2}\end{equation}
where 
\begin{equation}
\tilde f(u)=f'(u)-\frac{1}{4(u^2-\Lambda^4)}g(u)\ \ , \qquad
\tilde g(u)=f(u)+g'(u)\ \ .
\label{cicciopoldo}\end{equation}
It is now clear what the form of the solutions of 
Eq.(\ref{razzo}) is. In fact, requiring $\tilde f=0$, that is
\begin{equation}
f'(u)-\frac{1}{4(u^2-\Lambda^4)}g(u)=0\ \ ,
\label{eiuoh}\end{equation}
we get the key relation 
\begin{equation}
a_{D,nc}'=H(u)a_D'\ \ ,\qquad a'_{nc}=H(u)a'\ \ ,
\label{wodihw3}\end{equation}
where
\begin{equation}
H(u)= f+8uf'+4(u^2-\Lambda^4)f''\ \ .
\label{oiuce}\end{equation}
Summarizing, from (\ref{wodihw}) and (\ref{eiuoh}) we have  
\begin{equation}
a_{D,nc}=f(u)a_D+4(u^2-\Lambda^4)f'(u)a_D'\ \ ,\qquad
a_{nc}=f(u)a+4(u^2-\Lambda^4)f'(u)a'\ \ .
\label{wodihwn}\end{equation}
which satisfy (\ref{razzo}) since, from (\ref{wodihw3}) we see that
\begin{equation}
\tau_{nc}={a_{D,nc}'\over a_{nc}'}={H(u)a_{D}'\over H(u) a'}
=\tau\ \ .
\label{oijw}\end{equation}

Until now we have derived a set of solutions of Eq.(\ref{razzo}) 
depending on the function $f$. Comparing (\ref{pocjw}), 
(\ref{d0idj}) and (\ref{anfamone}) with (\ref{wodihwn}), we see that 
the function $f$ should satisfy the conditions
\begin{equation}
f(u\to\infty)=1\ \ ,\qquad f(u\to \Lambda^2)
\sim d_0(u-\Lambda^2)^{-2}\ \ .
\label{oifjhe}\end{equation}
Let us set
\begin{equation}
f(u)={P(u)\over Q(u)}\ \ .
\label{cijc}\end{equation}
$P$ and $Q$ should be polynomial functions, 
since otherwise we would get singularities not found in the 
asymptotic analysis. The first condition in (\ref{oifjhe})
fixes $P$ and $Q$ to be of the same degree, while from the second
condition we obtain 
\begin{equation}
Q(u)=(u-\Lambda^2)^2\sum_{k=0}^{N}c_k u^k\ \ .
\label{opfjwe2}\end{equation}

Due to the singularity structure, it is reasonable to assume that 
the only possible poles in the finite region of the moduli space 
arise at the punctures
$u=\pm\Lambda^2$. Another condition concerns the
$\Z_2$ symmetry of the moduli space. To understand this, let
us recall that, in the commutative case, $a_D(e^{i\pi/2}a)=a_D-a$ and
$a(-u)=e^{i\pi/2}a$ \cite{M3}. In order to preserve these
properties for
$a_{D,nc}$ and $a_{nc}$, we need 
that $P(-u)=P(u)$ and $Q(-u)=Q(u)$, so that
\begin{equation}
Q(u)=(u^2-\Lambda^4)^2\sum_{k=0}^{N-2}\tilde c_k u^k\ \ .
\label{opfjwe3}\end{equation}
Thus we end with an expression which is singular at 
$u=\pm\Lambda^2$. Concerning the coefficients $\tilde c_k$ we note 
that $\tilde c_{k\neq0}=0$, since otherwise we would have poles 
outside the critical points.

Summarizing, we have
\begin{equation}
f(u)={u^4+\alpha u^2+\beta\over (u^2-\Lambda^4)^2}\ \ ,
\label{opfjwe4}\end{equation}
where $\alpha$ and $\beta$ are functions of $\Lambda$ and of the 
noncommutative parameter $\theta$. Note that this 
implies that the constants $c_0$ and $d_0$ in (\ref{d0idj}) and 
(\ref{oifjhe}), are 
\begin{equation}
c_0=\frac{i}{2\Lambda}d_0\ \ ,\qquad
d_0=\Lambda^8+\alpha\Lambda^4+\beta\ \ .
\label{maddeche}\end{equation} 
There is still one 
more condition we have 
to satisfy. Namely, in the
$\theta\to0$ limit, $(a_{D,nc},a_{nc})$ should reduce to
$(a_D,a)$. This implies that $\lim_{\theta\to0}f=1$, that is
\begin{equation}
\lim_{\theta\to0}\alpha=-2\Lambda^4\ \ ,\qquad
\lim_{\theta\to0}\beta=\Lambda^8\ \ .
\label{oiwhdf2}\end{equation}
These conditions together with dimensional analysis imply
\begin{equation}
\alpha=\Lambda^4\left[-2+\sum_{k=1}^\infty \alpha_k 
(\theta \Lambda^2)^k\right]\ \ ,
\qquad\beta=\Lambda^8\left[1+\sum_{k=1}^\infty \beta_k (\theta
\Lambda^2)^k\right]\ \ .
\label{alfabeta}\end{equation}

Notice that the expressions of $a$ and $a_D$ get modified to
\begin{equation}
a_{D,nc}=2\int_{\Lambda^2}^u \lambda_{nc},\qquad 
a_{nc}=2\int_{-\Lambda^2}^{\Lambda^2}\lambda_{nc}\ \ ,
\label{ikudq2}\end{equation}
where, from (\ref{wodihwn})
\begin{equation}
\lambda_{nc}=f\lambda+4(u^2-\Lambda^4)f'\lambda'\ \ ,
\label{oioiaoimmeia}\end{equation}
where $\lambda$ stands for the Seiberg--Witten differential 
\begin{equation}
\lambda={\sqrt 2\over 2\pi}
{dx\sqrt{x-u}\over \sqrt{x^2-\Lambda^4}}\ \ ,
\label{saibeuitte}\end{equation}
Besides the divergence in the mass of the monopole found in
\cite{Yoshida}, we see that the BPS mass formula has divergences
both at $u=\Lambda^2$ and $u=-\Lambda^2$ for any nontrivial value of 
$n_e$ and $n_m$
\begin{equation}
M=\sqrt2|n_e a_{nc}+n_m a_{D,nc}|\ \ .
\label{odfij}\end{equation}

It is of great importance to investigate the structure of the
expansions for $\alpha$ and $\beta$. Their explicit form will 
determine the critical 
values of $\theta$, $n_e$ and $n_m$ corresponding to possible 
cancellations of
divergences and the appearance of possible zeros for $M$. Let us 
conclude by observing that, despite many technical difficulties, a 
noncommutative analogue 
\cite{NEKSCH} of the analysis of the instanton calculations
performed in the context of the standard 
Seiberg--Witten model \cite{DKM,BFMT} is relevant in 
order to fix the structure of $\alpha$ and $\beta$.

\acknowledgments{J.M.I. is supported by an INFN fellowship. 
J.M.I. and M.M. are partially supported by the 
European Commission TMR program ERBFMRX-CT96-0045.}


\begin{thebibliography}{99}

\bibitem{CONNES}
A. Connes, M. Douglas and A. Schwarz, JHEP {\bf 9802} (1998) 003.

\bibitem{SW2}
N. Seiberg and E. Witten, JHEP {\bf 9909} (1999) 032.

\bibitem{DH}
M.R. Douglas and C. Hull, JHEP {\bf 9802} (1998) 008.

\bibitem{SH}
M.M. Sheikh-Jabbari, Phys. Lett. {\bf B450} (1999) 119.

\bibitem{SHEIKH}
M.M. Sheikh-Jabbari, {\it Noncommutative Super Yang-Mills theories 
with 8 supercharges and brane configurations}, {\tt hep-th/0001089}.

\bibitem{Yoshida}
Y. Yoshida, {\it Nonperturbative aspect in $N=2$ supersymmetric 
noncommutative Yang-Mills theory}, {\tt hep-th/0009043}.

\bibitem{SW}
N. Seiberg and E. Witten, Nucl. Phys. {\bf B426} (1994) 19.

\bibitem{SEI}
M. Van Raamsdonk and N. Seiberg, JHEP {\bf 0003} (2000) 035; 
S.~Minwalla, M. Van Raamsdonk and N. Seiberg, {\it Noncommutative 
perturbative dynamics}, {\tt hep-th/9912072}.

\bibitem{M3}
G. Bonelli, M. Matone and M. Tonin, Phys. Rev. {\bf D55} (1997) 6466. 

\bibitem{M12}
M. Matone, Phys. Lett. {\bf B357} (1995) 342; Phys. Rev. {\bf D53} 
(1996) 7354; Phys. Rev. Lett. {\bf 78} (1997) 1412.

\bibitem{LERCHE}
A. Klemm, W. Lerche, S. Yankielowicz and S. Theisen, Phys. Lett.
{\bf B344} (1995) 169. 

\bibitem{NEKSCH}
N. Nekrasov and A. Schwarz, Commun. Math. Phys. {\bf 198} (1998) 689.

\bibitem{DKM}
D.Finnell and P. Pouliot, Nucl. Phys. {\bf B453} (1995) 225; N. 
Dorey, V.V. Khoze and M.P. Mattis, Phys. Rev. {\bf D54} (1996) 2921; 
Phys. Rev. {\bf D54} (1996) 7832; F. Fucito and G. Travaglini, 
Phys Rev. {\bf D55} (1997) 1099.

\bibitem{BFMT}
D. Bellisai, F. Fucito, M. Matone and G. Travaglini, Phys. Rev. {\bf 
D56} (1997) 5218. 



\end{thebibliography}
\end{document}